# A Unifying Hypothesis for the Conformational Change of Tubulin

Deborah Kuchnir Fygenson
*Department of Physics, University of California, Santa Barbara, CA 93106*

Microtubule dynamic instability arises from the hydrolysis of GTP bound to the β-monomer of the tubulin dimer. The conformational change induced by hydrolysis is unknown, but microtubules disassemble into protofilaments of GDP-bound tubulin that curve away from the microtubule axis. This paper presents the unfolding of a portion of the tubulin molecule into the microtubule interior as a plausible, unifying explanation for diverse structural and kinetic features of microtubules. This is the first specific structural hypothesis for the hydrolysis induced conformational change of tubulin that simultaneously explains weakening of lateral bonds, bending about longitudinal bonds, changes in protofilament supertwist associated with GTP hydrolysis, structural features of GDP-tubulin double rings, faster disassembly at higher temperatures and slower disassembly in the presence of glycerol and deuterium oxide. As such, the hypothesis makes a strong case for further theoretical investigation and direct experimental tests.

Microtubules are hollow, cylindrical aggregates of the protein tubulin, 25 nm in diameter. *In vivo* as well as *in vitro*, individual microtubules switch repeatedly between growing (assembling) and shortening (disassembling) from their ends, exhibiting macroscopic fluctuations in overall length ($\Delta \ell \sim \ell$) under otherwise constant chemical conditions (Kirschner and Mitchison, 1986). This puzzling "dynamic instability" is fueled by the hydrolysis of tubulin-bound guanosine tri-phosphate (GTP) (Hyman et al., 1992), which is loosely understood as triggering a conformational change in the tubulin molecule that eventually destabilizes the aggregate (Tran et al., 1997). The structure of tubulin has been solved with 3.7Å resolution (Nogales et al., 1998b), but the physical nature of the conformational change is not yet known (Downing and Nogales, 1998b).

This paper presents and explores the hypothesis that the hydrolysis-induced conformational change of tubulin involves a portion of the sparsely structured N-terminal domain of the β-tubulin molecule "unfolding" into the interior of the microtubule. It is shown that such a change can destabilize the aggregate in a manner consistent with structural data (Mandelkow et al., 1991). From the data, the effective size of the unfolded region is deduced and a candidate locus in the three dimensional structure of tubulin is identified. The hypothesis leads to novel interpretations of two important structural results: the hydrolysis associated change in microtubule supertwist (Hyman et al., 1995), and the structure of GDP-tubulin double-rings (Diaz et al., 1994). It also provides a unifying explanation for the effects of temperature (Fygenson et al., 1994) and glycerol (Fygenson, 1995) on microtubule disassembly rates.

These results motivate further theoretical and experimental investigations. If verified, the existence of localized unfolding will enable major advancements in the understanding of microtubules. It will also establish the biological relevance of localized unfolding as a subset of protein conformational change, opening the way for quantitative modeling and testing of a novel class of molecular machines.

## BACKGROUND

Tubulin is a heterodimer of nearly identical α- and β-tubulin polypeptides (Burns and Surridge, 1994). These monomers each bind one molecule of GTP and are bound to one another in a head-to-tail arrangement (αβ) with the GTP-binding site of α-tubulin buried at the interface and the GTP-binding site of β-tubulin exposed at the opposite pole (Nogales et al., 1998b). Dimers spontaneously aggregate head-to-tail into protofilaments, burying the β-bound GTP at the dimer-dimer (αβ-αβ) interface. Protofilaments aggregate in parallel (ββ,αα) to form the microtubule wall. Lateral bonds are slightly tilted and angled inwards, so that 13 protofilaments close naturally to form a tube whose axis is parallel to the protofilament axis and whose surface lattice is described by a left-handed three-start helix, with a seam (where the tube closed) along which monomers of one type interface laterally with monomers of the other type (Mandelkow et al., 1986). (See Fig. 1).

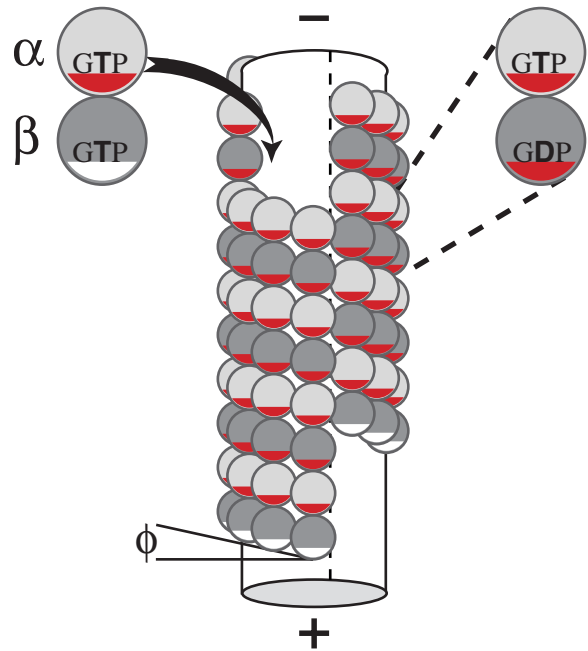

**Fig. 1.** Schematic of microtubule structure. In solution, tubulin binds GTP non-exchangeably (red) on the α-monomer and exchangeably (white) on the β-monomer. The GTP-laden dimer aggregates into a left-handed, three-start helical lattice characterized by an angle $\phi$. Microtubules typically form with 13 protofilaments that align parallel to the axis of the tube. Monomers interface laterally with monomers of the same type, except along one protofilament pair that forms a seam (dashed line). Within the microtubule, GTP bound to the β-monomer is hydrolyzed to GDP, which is non-exchangeable (red) until the dimer returns into solution after disassembly.

The main difference between α- and β-tubulins is in binding GTP. α-bound GTP is effectively sequestered – not exchanged and not hydrolyzed – but β-bound GTP is labile – exchangeable in the free dimer and hydrolyzed to (non-exchangeable) guanosine di-phosphate (GDP) in the protofilament (Weisenberg et al., 1976). A significant amount of the free energy of this hydrolysis goes into the microtubule via a conformational change of the tubulin dimer (Caplow et al., 1994).

Although the hydrolysis reaction is closely coupled to microtubule assembly (Carlier and Pantaloni, 1981; Stewart et al., 1990), its consequence is to destabilize the structure. Experiments indicate that unhydrolyzed GTP-tubulin is limited to the last layer of subunits at the end of a microtubule (Voter et al., 1991; Walker et al., 1991; Drechsel and Kirschner, 1994). The usual interpretation is that this layer acts as a "GTP-cap", keeping an otherwise unstable microtubule intact (Mitchison and Kirschner, 1984). Numerical and analytical models along these lines reproduce some salient features (Martin et al., 1993; Flyvbjerg et al., 1996) but do not imply a microscopic mechanism to justify the tuning of their parameters. At this point, an understanding of microtubule dynamics awaits greater knowledge of the hydrolysis-induced conformational change (Downing and Nogales, 1998a).

Circular dichroism (Howard and Timasheff, 1986) and Raman spectroscopy (Audenaert et al., 1989) show a slight change in tubulin secondary structure content after hydrolysis, but by far the strongest signature of the conformational change has appeared in electron micrographs of growing and shortening microtubules (Mandelkow et al., 1991). At the ends of growing microtubules, protofilaments of different lengths are straight and closely associated. In shortening microtubules, protofilaments separate from one another and curl back, away from the microtubule axis, forming characteristic blunt "blossoms" at the microtubule ends (Tran et al., 1997; Muller-Reichert et al., 1998). This observation is the basis for comments about the tubulin dimer adopting a "curved" or "kidney-bean shaped" conformation, held under tension in the microtubule lattice (reviewed in (Tran et al., 1997; Downing and Nogales, 1998a)). Localization of the hydrolyzed nucleotide at the inter-dimer interface is highly suggestive, but which and how residues shift to generate tension and weaken lateral bonds is an important subject of active research (Davis et al., 1994; Sage et al., 1995).

## HYPOTHESIS

Consider α- and β-tubulin as solid spheres of radius $b$. Suppose that hydrolysis of β-bound GTP causes a portion of β-tubulin, initially condensed onto the monomer surface, to "unfold" into solution. Over time, this domain will explore a volume of radius $\rho$ centered on the surface of the β-tubulin sphere (some of which is excluded by the sphere itself) (Fig. 2a)., This unfolded region is herein referred to as an *entropic bristle domain* (EBD) to evoke its analogy with the entropic brushes commonly used to inhibit aggregation in colloids (Hoh, 1998; Milner, 1991).

If they are tethered to the side of β-tubulin that faces into the microtubule interior, EBDs will be sterically hindered.

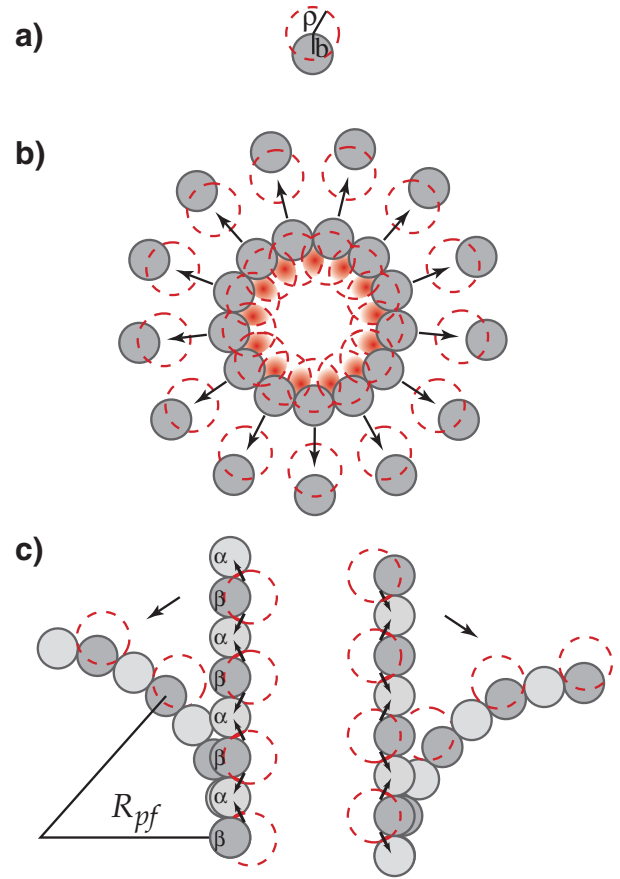

**Fig. 2.** Schematics of the proposed entropic bristle domain (EBD) on β-tubulin. **a)** Side view of a single β-tubulin monomer (gray) of radius $b$, with an EBD (red) of characteristic size $\rho$. **b)** Cross-section of a microtubule cut along a row of β-tubulins. Shading (red) indicates regions of overlap where EBDs sterically hinder one another. Radial expansion of the aggregate releases EBDs from confinement at the expense of lateral bonds. **c)** Side view of a pair of protofilaments oriented as in a microtubule. For simplicity, the α-tubulin monomers (light gray) are drawn contacting the β-tubulin monomers (dark gray) equidistant on either side of the EBD (red). Note the slight overlap of the EBDs with the α-tubulin spheres. A radius of curvature $R_{pf}$ eliminates the overlap.

In the microtubule surface lattice, a row of β-tubulins follows a helix of radius $R_\mu$ with a pitch given by the angle $\phi$ (Fig. 1). Geometry dictates that EBDs on laterally adjacent β-tubulins will overlap if

$$\rho > b\Big(1 - \big(b\cos\phi\big)/R_\mu\Big).$$

In the longitudinal direction, by contrast, β-tubulins are separated by α-tubulins (Fig. 1). α-tubulins do not hydrolyze GTP and so presumably do not release EBDs. Thus, longitudinally spaced EBDs will not overlap unless $\rho > 2b$. They will be sterically hindered, nevertheless, by overlap with neighboring α-tubulin "spheres". If the EBD anchored on a β-tubulin is at an angle $\xi$ from a neighboring α-tubulin, the geometrical criterion for overlap can be written as

$$\rho \geq b\Big(\sqrt{5 - 4\cos\xi} - 1\Big).$$

The criteria for overlap are important because the resulting steric hindrance between EBDs costs entropic free energy. The cost is greater in the lateral direction, where it is equivalent to a pressure radially outward that favors separation of the protofilaments (Fig. 2b). In the longitudinal direction, there is less overlap, but still there is pressure on the neighboring α-tubulins that is relieved by increasing ξ. Thus, a straight protofilament (ξ=90°) will, upon the unfolding of EBDs, curve back and away from the microtubule axis (ξ>90°) as it separates from its neighbors (Fig. 2c).

In this way, a conformational change consisting of nothing more than the release of an EBD into the microtubule lumen can explain the most striking structural characteristics of a disassembling microtubule (Mandelkow et al., 1991; Tran et al., 1997). Preferential breaking of lateral bonds over longitudinal ones is expected, even if the two are comparable in strength, because lateral bonds are subject to greater stress (from closer packed EBDs) than longitudinal ones. Curling of separated protofilaments is expected, even if α-tubulins do not change conformation, because curvature places α-tubulins outside the EBDs on neighboring β-tubulins.

### Size of the unfolded domain

The effective size of the hypothetical EBD on β-tubulin can be deduced from measurements (Mandelkow et al., 1991) of the radius of protofilament curvature, $R_{pf}$, using a geometrical no-overlap criterion

$$\rho = b\left(\sqrt{5 + 4b/R_{pf}} - 1\right).$$

For $R_p = 19$ nm and $b = 2$ nm, this equation gives $\rho = 2.7$ nm. (Note: This result assumes the EBD is anchored equidistant from the two neighboring α-tubulins along the protofilament, but the calculation is relatively insensitive to the location of the anchor point. Anchoring closer to one α-tubulin than the other gives $\rho < 2.9$ nm.)

Given this size, it is possible to guess roughly the number of amino acids involved in the EBD. If $\rho$ is interpreted (albeit loosely) as the radius of gyration of the self avoiding random walk executed by backbone of the EBD polypeptide, then it should scale (Doi and Edwards, 1986) with the step size, $a$, and the number of steps, N, to the 3/5 power, $\rho \approx N^{3/5} a$. Taking $\rho=2.7$nm and $a=3$Å this relation suggests that the EBD might involve ~40 amino acids.

### Location of the unfolded domain

The structure of tubulin in an assembly competent, taxotere stabilized conformation is known with 3.7Å resolution (PDB entry TUB1) (Nogales et al., 1998b). The authors of that work point out that the N-terminal domain contains "…long predicted loops on the putative inside surface of the microtubule [that] correspond to a region of the sequence that can accommodate insertions and deletions." In particular, after the first β-strand (B1; residues 3-8) and α-helix (H1; residues 12-23), there is a large loop (residues 24-64) before a short β-strand (B2; residues 65-70). The next, rather short, α-helix (H2; residues 74-79) is also followed by a loop (residues 80-92) and a short β-strand (B3; residues 93-97). Yet another large loop (residues 97-110) precedes the third α-helix (H3; residues 111-127). (See Fig. 3.)

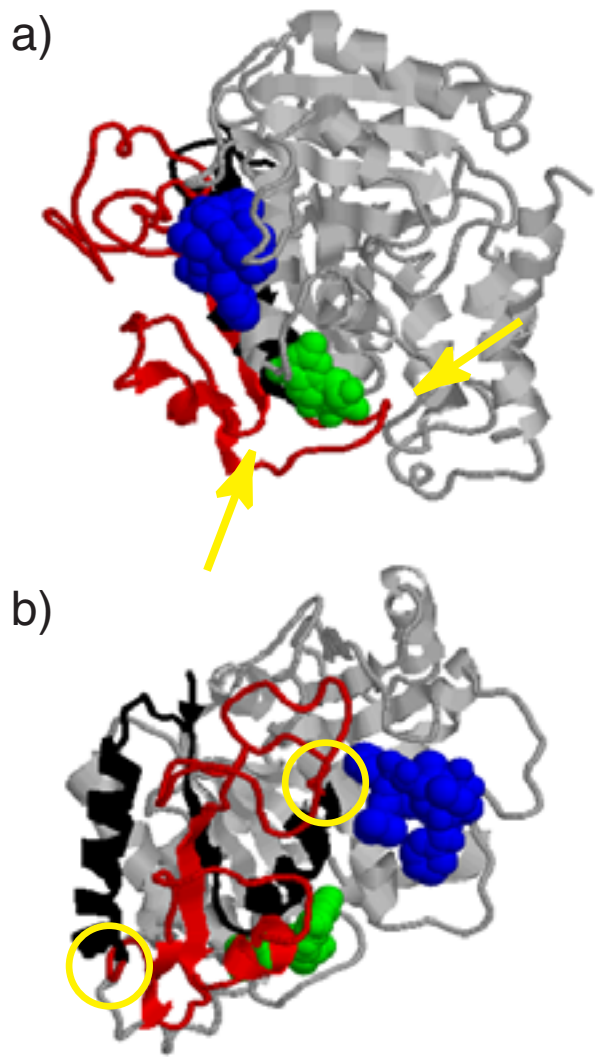

**Fig. 3.** Cartoon structure of β-tubulin, from entry 1TUB in the Protein Data Bank (http://www.rcsb.org/pdb/) using RasMol (http://www.umass.edu/microbio/rasmol/getras.htm). **a)** view in profile: microtubule interior on the left and exterior on the right. **b)** view from the center of the microtubule. The supposed EBD is colored red. Portions of the N-terminal domain which might anchor the EBD to the protein core are colored black (residues 1-23, 111-133). The GDP-ligand is green and the taxotere molecule is blue. Arrows indicate residues thought to interact with the β-bound GTP. Circles indicate potential anchor sites.

Although they appear close packed in the space-filling rendition of the published crystal structure, the authors specifically note, "Strands B2 and B3 are poorly defined in the unrefined density map. The residues in the loops connecting H1 and B2, and H2 and B3, are included for completeness, but were built in very weak density."

Thus, the N-terminal domain of β-tubulin is a plausible locus for the hypothetical EBD. Although they wrap around themselves, the loops in this region (residues 24-110) are not intertwined with the rest of the molecule (Fig. 3a). In any rendition, the EBD would involve at least the first, large loop (residues 24-64) of the N-terminal domain. One possibility is that the EBD is a ~80 amino acid loop (residues 24-110) anchored to the β-tubulin surface at its two ends. In the N-terminal domain, at least three segments appear to interact with the GDP (Nogales et al., 1998a), so some residues in the loops may well be held near the β-tubulin surface by a third phosphate group. Furthermore, a stretch near the top of H1 (residues 15-25) interacts with the taxotere molecule (Fig. 3b), suggesting that this drug's stabilizing effect might derive from an attractive interaction that effectively localizes the EBD. This would explain the otherwise surprising location of the drug binding site on the microtubule interior (Amos and Lowe, 1999).

## IMPLICATIONS FOR STRUCTURES

The estimated size, $\rho$, of the hypothetical EBD is particularly interesting in relation to the spacing of EBDs across the seam in the microtubule. The dimensions are such that

$$2b^2\left(1 - b/R_\mu\right)^2 < \rho^2 < 2b^2.$$

The left-hand inequality implies that EBDs overlap across the seam when curved toward each other in a microtubule. The right hand inequality implies that EBDs would not overlap across the seam if not constrained to curve toward each other, as might occur when a microtubule disassembles. The rest of this section explores two consequences of these two inequalities that lead to new interpretations of structural data.

### Slipping at the seam

Although the apparent size of the EBDs should be large enough to repel across the seam inside a microtubule, the microtubule surface lattice is not square, so the repulsion would not be symmetric across the seam (Fig. 4a). To balance the forces, the mismatched protofilaments should slip so as to *increase* the left-handed helical pitch of the monomers. The distance slipped, $\Delta$, will cause the protofilaments to adopt an angle $\theta$ with respect to the microtubule axis such that $\tan\theta = \Delta/2bN$, where $N$ is the number of protofilaments and $2b$ is the width of a protofilament. The angle means that the protofilaments twist around the microtubule with a characteristic pitch, $L_p$, given by

$$L_p = \frac{2bN\cos\theta}{\tan\theta} = \frac{(2bN)^2}{\Delta}\cos\theta \approx \frac{(2bN)^2}{\Delta}$$

where the approximation holds for small $\theta$ (i.e., $\Delta \ll 2bN$).

Twisting protofilaments are well known in microtubules which form with a protofilament number different from 13 (Wade et al., 1990; Chretien and Wade, 1991). In such microtubules, protofilaments at the seam must "shift" a distance $\varepsilon$ to make lateral bonds come into register. The resulting "supertwist" has a pitch $L_p \approx (2bN)^2/\varepsilon$ that is, by convention, negative when it is left-handed (i.e., $\varepsilon < 0$ for shifts that go decrease the three-start helical pitch) and positive when it is right-handed (i.e., $\varepsilon > 0$ for shifts that increase the three-start helical pitch). Thus, the release of EBDs should cause protofilaments to develop a tighter right-handed or looser left-handed supertwist, according to

$$L_p \approx (2bN)^2/(\varepsilon + \Delta).$$

This argument suggests a new interpretation of cryo-electron microscopy data that measured the protofilament supertwist in microtubules "before" and "after" hydrolysis (Hyman et al., 1995). For the former, microtubules were assembled from tubulin bound to GMPCPP, a slowly hydrolyzable analogue of GTP. For the latter, microtubules were assembled in the standard way, in the presence of GTP, which quickly hydrolyzes. Measurements were made on microtubules with $N$=12, 13, 14, 15.

Before hydrolysis, microtubules with $N$=14, 15 had a pronounced right-handed supertwist ($L_p$=9.0, 4.1µm, respectively) whereas microtubules with $N$=13 showed a slight left-handed supertwist ($L_p$=-14.8µm). The supertwist handedness of microtubules with $N$=12 ($|L_p|$=3.7 µm) was not reported. After hydrolysis, the supertwist pitch was different: tighter for the right-handed ($\varepsilon$>0) microtubules ($L_p^*(14, 15)$=6.6, 3.5µm) and looser for the left-handed ($\varepsilon$<0), ($L_p^*(N=13)$=∞) and undetermined microtubules, ($|L_p^*(N=12)|$=4.7µm).

From this data, the EBD hypothesis predicts that the microtubules with $N$=12 were left-handed. Furthermore, using the first set of measurements (before hydrolysis, $\Delta$=0) to estimate $\varepsilon$, the EBD hypothesis allows one to derive from the second set of measurements (after hydrolysis, $\Delta$>0) estimates of $\Delta$=1.3, 1.8, 1.3, 1.5 Å for respective values of $N$=12, 13, 14, 15. These values are consistent with a 1.5 Å shift in layer-line spacing reported in the same experiment and provide a coherent interpretation of the data for all values of N. The alternate interpretation of the 1.5Å shift in layer-line spacing as an overall shortening of the tubulin dimer is not consistent with the looser left-handed supertwists reported for $N$=12 and $N$=13 microtubules (Hyman et al., 1995).

### Spirals from the seam.

Given the structure of the microtubule surface lattice, EBDs on either side of the seam should not overlap as much as other neighboring EBD pairs (Fig. 4a). In fact, given their size, EBDs on either side of the seam might not overlap at all if the protofilaments were not constrained to angle inward (i.e., $R_\mu \to \infty$). The EBD hypothesis therefore suggests that, upon disassembly, the protofilaments that comprise the seam might not separate from one another. Instead, they could curl back and away from the microtubule axis as a pair, just like a single protofilament. If the protofilaments in the pair rotate slightly about their long axes (Fig. 4a, inset), their EBDs would avoid all steric hindrance, even that due to overlap with laterally adjacent α-tubulins. The overlap condition

$$\rho = b\left(\sqrt{5 + 4\cos\phi\sin\delta} - 1\right),$$

with $\rho$=2.7nm, $b$=2nm and $\phi$<20°, indicates that a rotation of $\delta$=8° would suffice. The total bend angle between laterally adjacent monomers, $2\delta$=16°, would then be similar to the bend in the longitudinal bond between monomers along a curled protofilament (Howard and Timasheff, 1986; Melki et al., 1989).

Furthermore, if an isolated seam were to curl straight back onto itself, it could form a stable spiral, with the outer protofilament pair nested between EBDs on the inner one (Fig. 4b). EBDs on the inner pair would splay further to each side as they press against the outer pair (Fig. 4b, inset) and the distance separating the protofilament cores, $x$, would be related to the angle of rotation $\gamma$ of the inner EBDs according to

$$(\rho + b)^2 = x^2 + b^2 - 2xb\cos\gamma.$$

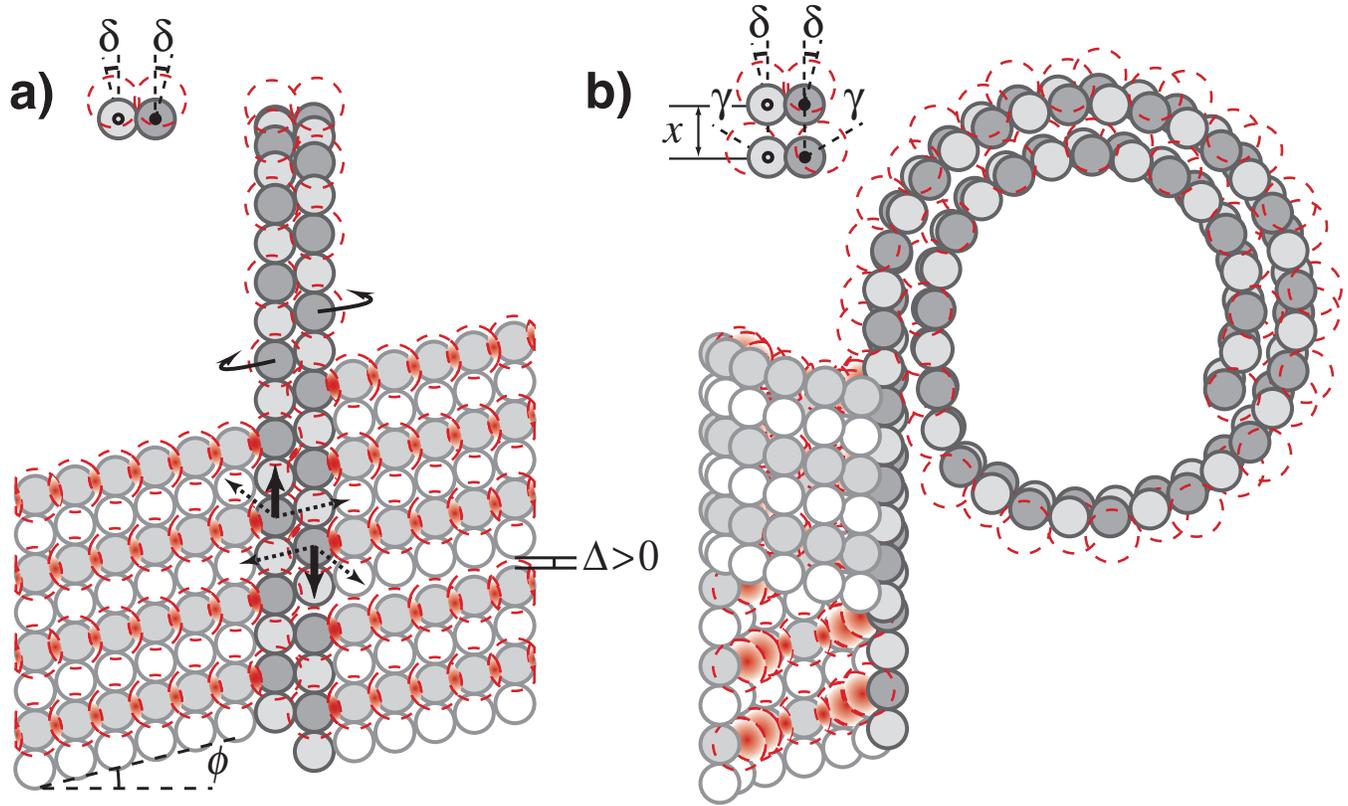

**Fig. 4.** Schematics of the microtubule lattice with emphasis on the seam. α-tubulins are light, β-tubulins are dark, EBDs are outlined in red dashes and regions of overlap are shaded in red. **a)** Microtubule lattice cut open, laid flat and viewed from the inside. The helical pitch of the lattice is characterized by the angle $\phi$ between a row of monomers and the horizontal. The upper portion depicts a perfectly rhombic lattice, as would be expected in the absence of EBDs. Dashed arrows represent forces acting on β-tubulins at the seam due to overlapping EBDs. Solid arrows show resultant forces which cause slippage at the seam. (The lower portion depicts a lattice with a defect along the seam, caused by slipping a distance $\Delta$.) **Top**: During disassembly, protofilaments at the seam curve back and away from the microtubule axis without separating. The isolated pair could adopt the preferred rhombic lattice motif while rotating slightly apart to eliminate any hindrance of the EBDs (curved arrows). The angle of rotation, $\delta$, for each protofilament, as viewed along the protofilament axis, is defined in the inset at left. **b)** Microtubule lattice closed into a tube, with an isolated seam, viewed in profile, curling back upon itself into a spiral. The EBDs on the inner protofilament pair are pressed against the outer protofilament pair. The angles of rotation, $\delta$ and $\gamma$, for protofilament pairs in the outer and inner rings, respectively, and the separation between the pairs, $x$, viewed along the protofilament axis, are defined in the inset.

This provides an alternate interpretation of the structure of the so-called double-rings of GDP-tubulin, often seen among the disassembly products of microtubules (Howard and Timasheff, 1986; Melki et al., 1989). The best resolved measurements of GDP-tubulin double-rings, from x-ray scattering of concentrated solutions (Diaz et al., 1994), can not differentiate between α and β tubulins, but do resolve the separation between the rings, x=5.5nm, and the dimensions and orientations of the ellipsoidal repeat unit: 4nm (tangent to the ring) × 7nm (at -60° from the radial axis) × 8nm (perpendicular to the plane of the ring). As a spiraled seam, the observed inter-ring separation, $x$=5.5nm, and the estimated EBD size, $\rho$=2.7nm, would together explain the size (7nm~$2b+\rho$) and orientation, $\gamma$=56°, of the intermediate ellipsoidal axis. Similarly, the major axis (8nm=$2b+2b$) would be interpreted as crossing the seam and the minor axis (4nm=$2b$) as the size of a single tubulin monomer along a protofilament.

A spiraled seam leads to a very different interpretation of the double-ring structure than suggested by the authors of the x-ray scattering study (Diaz et al., 1994). They propose that the double rings are only one monomer thick, and "correspond to two neighbor protofilaments curved" in a direction "tangent to the microtubule surface (i.e., sideways)". In addition to the unusual monomer thickness, this interpretation is puzzling, as the authors note, in that "lateral contacts between monomers in the microtubule become necessarily out of register in the outer and inner rings with different numbers of subunits". Lateral bonds between tubulins, suggested in the sideways ring structure, would usually require a specific alignment of their surfaces. No particular positioning between tubulins on the two rings would be necessary, however, for a pair of rings derived from a spiraled seam with EBDs. Steric interactions would hold the outer ring between EBDs on the inner ring.

The notion that double rings derive from a spiraled seam structure is supported by two cryo-electron microscopy studies (Mandelkow et al., 1991; Nicholson et al., 1999). In the earlier study, cross sectional views of nested rings composed of protofilament pairs are explicitly documented. (See Fig. 3 in (Mandelkow et al., 1991).) In the more recent study, reconstructions averaging under imposed rotational symmetry clearly show that the outer ring has been curved "so that the 'bumpy' side of tubulin, which in the microtubule corresponds to the inside surface, now faces the outside of the double ring, whereas the flatter side faces the inside." (See Fig. 4. in (Nicholson et al., 1999)).

## IMPLICATIONS FOR DISASSEMBLY

Another consequence of the EBD hypothesis is that microtubule disassembly is driven, at least in part, by the entropic free energy gained when a cohort of unfolded domains escape the confinement of the microtubule lumen. This leads to qualitative predictions about the microtubule disassembly rate as a function of intensive variables such as temperature, concentration of glycerol and concentration of deuterium oxide ($D_2O$). The rest of this section compares predictions with experiments using these three variables.

### Effect of temperature

Because the free energy attributable to EBD overlap is inherently entropic, it should increase at higher temperatures and lead to faster rates of microtubule disassembly. At first this prediction appears to contradict the well known fact that microtubules are, overall, more stable at higher temperatures (Dustin, 1984). However, measurements of single microtubule dynamics as a function of temperature show that both assembly *and disassembly* of individual microtubules are faster at higher temperatures (Fygenson et al., 1994). The greater stability overall is due to a reduction in the frequency of transitions from growth to shortening (catastrophe) and an increased probability of the reverse (rescue). The increased rate of assembly indicates a strengthening of lateral bonds at higher temperatures that, by itself, would lead to a decreased rate of disassembly. The EBD hypothesis thus provides an elegant and plausible interpretation of the increased disassembly rate.

### Effect of glycerol and $D_2O$

Because the size of an EBD is effectively its time-averaged volume in the surrounding solvent, poor solvents should decrease $\rho$, reduce the amount of overlap, and lead to slower rates of microtubule disassembly. Glycerol and $D_2O$ are standard "poor solvents" that generally reduce the specific volume of proteins (Kresheck et al., 1965; Timasheff, 1993; Priev et al., 1996). Both are known to increase the extent and stability of microtubules in bulk (Schilstra et al., 1991; Chakrabarti et al., 1999). A common explanation is that the poor solvents increase the strength of hydrophobic interactions between tubulin dimers. However, measurements of single microtubule assembly and disassembly require a different explanation.

Increasing the concentration of glycerol up to 3M (Fygenson, 1995) has no effect on the rate at which single microtubules assemble, but causes a steady decrease in their rate of disassembly (Fig. 5). Increasing the concentration of $D_2O$ up to 60% does the same (see Fig. 4 in (Panda et al., 2000)). The constant assembly rate indicates that the free energy in lateral bonds is unchanged by the co-solvents. Therefore, the increased rate of disassembly must be attributed to the nature of the hydrolysis-induced conformational change and is, again, consistent with the EBD hypothesis.

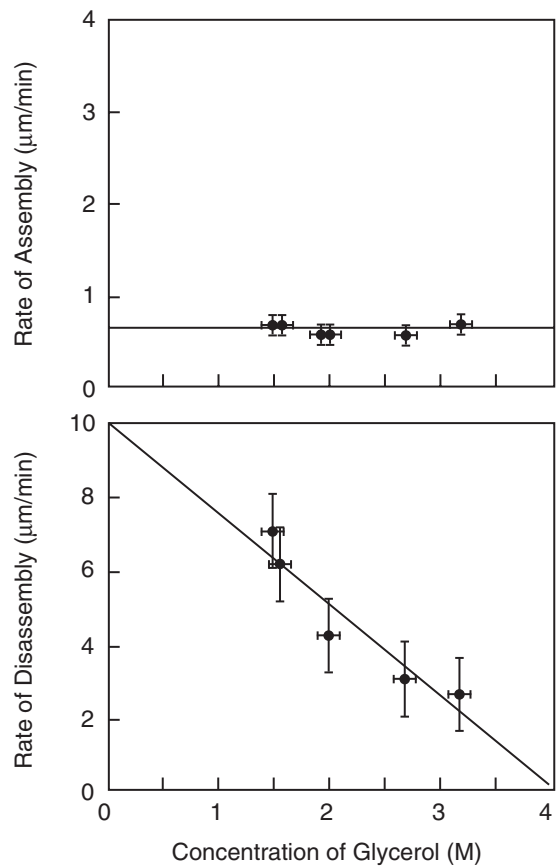

**Fig. 5.** Effect of glycerol on the average rate of **a)** assembly and **b)** disassembly at the plus ends of individual microtubule (reprinted from (Fygenson, 1995)). Methods are as described in (Fygenson et al., 1994). Tubulin concentration was 14 µM in buffer containing 1mM GTP, 2mM $MgSO_4$, 2mM EGTA, and 100mM Pipes at a pH 6.9 and temperature 18.5°C. Points are averages from a total of 1 hour of observation. Error bars represent one standard deviation. In a) the line represents the assembly rate in the absence of glycerol. In b) the line is a least-squares fit to the data.

## CONCLUSION

The hypothesis presented here, in which a portion of the β-tubulin monomer unfolds into an entropic bristle domain on the interior of the microtubule, provides a unifying explanation for several diverse results in the microtubule field, ranging from the structural to the kinetic. This hypothesis is significantly more powerful than current, practically unfalsifiable statements that describe GTP-hydrolysis as selectively weakening lateral bonds more than longitudinal ones and the tubulin dimer as consequently adopting a "kinked", "curved" or "kidney bean" conformation. The putative EBD is physically tractable, structurally specific, and highly suggestive of further calculations and experimentation that would refine and definitively test the hypothesis. For example, a calculation of the free energy cost of confining polypeptide loops on an asymmetric, inwardly curved lattice has the prospect of quantitative comparison with measurements of the free energy stored in lateral inter-tubulin bonds (Vulevic and Correia, 1997) and with changes in the rate of microtubule disassembly in different solvent conditions (Schilstra et al., 1991; Panda et al., 2000) and is the focus of ongoing research. A stability analysis of how such a quasi-one-dimensional polypeptide brush responds to edge-perturbations might lead naturally to a mechanistic model of the mysterious catastrophe and rescue transitions that characterize microtubule dynamic instability.

A definitive test of the EBD hypothesis would involve site-directed mutagenesis of tubulin (Davis et al., 1993) to change the net length of the disordered loops found in the N-terminal domain. Cryo-electron microscopy could be used to look for a commensurate change in the radius of curvature of protofilaments made of the purified mutant protein. Video-enhanced DIC microscopy could be used to quantify any associated change in the rate of single microtubule disassembly.

The EBD hypothesis is especially rich in corollary predictions that bear directly on open questions in the microtubule field. One example is the origin of sudden, discrete changes in assembly and disassembly rates of single microtubules (Gildersleeve et al., 1992). It has been suggested (Guenette and Solomon, 1993) that these changes are due to the presence of defects that change the microtubule protofilament number (Chretien et al., 1992). The EBD hypothesis predicts that microtubules with a greater number of protofilaments disassemble more slowly than microtubules with fewer protofilaments. This is because the volume in the microtubule lumen scales as the square of the number of protofilaments ($V \propto N^2$), whereas the number of EBDs scales only linearly ($n \propto N$), so the free energy cost of confinement, which increases with density, decreases with protofilament number ($F=f(n/V)=f(N/N^2)=f(N^{-1})$). The prediction could be tested by comparing disassembly rates of microtubules with different protofilament numbers.

Another example comes from microtubule protofilament number distributions. It is known that buffer conditions (Ray et al., 1993) or drugs (Andreu et al., 1994; Diaz et al., 1998) can influence the distribution of protofilament numbers in spontaneously nucleated microtubules, but little is known about why. The EBD hypothesis suggests a potentially insightful experiment, reasoning as follows: In the sheet-like tubulin aggregates that precede microtubule nucleation (Fygenson et al., 1995), hydrolysis occurs (Carlier et al., 1997) and some EBDs may be released before the sheets close to form tubes. The presence of EBDs would induce the sub-critical aggregates to adopt a larger radius of curvature and bias them toward closing with a greater number of protofilaments. The prediction is that conditions that slow or inhibit the release of EBDs (i.e., conditions that stabilize microtubules against disassembly) will nucleate microtubules with lower protofilament numbers. To observe the effect, it may suffice to simply tip the balance of time scales toward faster tubulin association as compared with hydrolysis and EBD release. If so, microtubules nucleated by rapid quench (high tubulin concentrations) would average fewer protofilaments than those nucleated near onset (low tubulin concentrations).

Yet another example is the question of mechanism underlying effects of microtubule-associated proteins and microtubule-binding drugs, such as taxol (Wilson et al., 1999). Agents which promote microtubule assembly but inhibit dynamics might be suspected to prevent the release of the EBD. Those which prevent assembly but do not induce disassembly could be involved in attracting the released EBD to another part of the tubulin surface.

It is my hope that the EBD hypothesis will lead to a stimulating dialogue between experiment and theory in the microtubule field. If this conformational change is eventually proven in tubulin, there will be significant impact outside the field as well. It would substantiate the notion that localized unfolding can be a biologically relevant type of conformational change (Hoh, 1998) and provoke similar hypotheses in other biological systems. And, because the energetics associated with EBDs are accessible to generalized physical theory and modeling, refinement of those theories in biological systems would open a path for the rational design of conformational change in artificial proteins or synthetic molecular machines.

I thank Drs. J.M. Andreu, H.P. Erickson, J. Hoh and S. Inoué for stimulating comments on an early version of the manuscript.

This work was supported in part by the MRL Program of the National Science Foundation under Award No. DMR00-80034 and by a National Science Foundation CAREER Grant No. 9985493.

**REFERENCES**

Amos, L. A., and J. Lowe. 1999. How Taxol stabilises microtubule structure. *Chem Biol.* 6:R65-69.

Andreu, J. M., J. F. Diaz, R. Gil, J. M. de Pereda, M. Garcia de Lacoba, V. Peyrot, C. Briand, E. Towns-Andrews, and J. Bordas. 1994. Solution structure of Taxotere-induced microtubules to 3-nm resolution. The change in protofilament number is linked to the binding of the taxol side chain. *J Biol Chem.* 269:31785-31792.

Audenaert, R., L. Heremans, K. Heremans, and Y. Engelborghs. 1989. Secondary structure analysis of tubulin and microtubules with Raman spectroscopy. *Biochim Biophys Acta.* 996:110-115.

Burns, R. G., and C. D. Surridge. 1994. . *In* Microtubules. J. S. Hyams and C.W.Lloyd, editors. Wiley, New York. 3-32.

Caplow, M., R. L. Ruhlen, and J. Shanks. 1994. The free energy for hydrolysis of a microtubule-bound nucleotide triphosphate is near zero: all of the free energy for hydrolysis is stored in the microtubule lattice [published erratum appears in J Cell Biol 1995 Apr;129(2):549]. *J Cell Biol.* 127:779-788.

Carlier, M. F., D. Didry, and D. Pantaloni. 1997. Hydrolysis of GTP associated with the formation of tubulin oligomers is involved in microtubule nucleation. *Biophys J.* 73:418-427.

Carlier, M. F., and D. Pantaloni. 1981. Kinetic analysis of guanosine 5'-triphosphate hydrolysis associated with tubulin polymerization. *Biochemistry.* 20:1918-1924.

Chakrabarti, G., S. Kim, M. L. Gupta, Jr., J. S. Barton, and R. H. Himes. 1999. Stabilization of tubulin by deuterium oxide. *Biochemistry.* 38:3067-3072.

Chretien, D., F. Metoz, F. Verde, E. Karsenti, and R. H. Wade. 1992. Lattice defects in microtubules: protofilament numbers vary within individual microtubules. *J Cell Biol.* 117:1031-1040.

Chretien, D., and R. H. Wade. 1991. New data on the microtubule surface lattice [published erratum appears in Biol Cell 1991;72(3):284]. *Biol Cell.* 71:161-174.

Davis, A., C. R. Sage, C. A. Dougherty, and K. W. Farrell. 1994. Microtubule dynamics modulated by guanosine triphosphate hydrolysis activity of beta-tubulin. *Science.* 264:839-842.

Davis, A., C. R. Sage, L. Wilson, and K. W. Farrell. 1993. Purification and biochemical characterization of tubulin from the budding yeast Saccharomyces cerevisiae. *Biochemistry.* 32:8823-8835.

Diaz, J. F., E. Pantos, J. Bordas, and J. M. Andreu. 1994. Solution structure of GDP-tubulin double rings to 3 nm resolution and comparison with microtubules. *J Mol Biol.* 238:214-225.

Diaz, J. F., J. M. Valpuesta, P. Chacon, G. Diakun, and J. M. Andreu. 1998. Changes in microtubule protofilament number induced by Taxol binding to an easily accessible site. Internal microtubule dynamics. *J Biol Chem.* 273:33803-33810.

Doi, M., and S. F. Edwards. 1986. The Theory of Polymer Dynamics. Oxford University Press, New York.

Downing, K. H., and E. Nogales. 1998a. Tubulin and microtubule structure. *Curr Opin Cell Biol.* 10:16-22.

Downing, K. H., and E. Nogales. 1998b. Tubulin structure: insights into microtubule properties and functions. *Curr Opin Struct Biol.* 8:785-791.

Drechsel, D. N., and M. W. Kirschner. 1994. The minimum GTP cap required to stabilize microtubules [published erratum appears in Curr Biol 1995 Feb 1;5(2):215]. *Curr Biol.* 4:1053-1061.

Dustin, P. 1984. Microtubules. Springer-Verlag, Berlin.

Flyvbjerg, H., T. E. Holy, and S. Leibler. 1996. Microtubule dynamics: Caps, catastrophes, and coupled hydrolysis. *Physical Review. E. Statistical Physics, Plasmas, Fluids, and Related Interdisciplinary Topics.* 54:5538-5560.

Fygenson, D. K. 1995. Microtubules: the rhythym of growth and evolution of form. Ph.D. Princeton University, Princeton.

Fygenson, D. K., E. Braun, and A. Libchaber. 1994. Phase diagram of microtubules. *Physical Review. E. Statistical Physics, Plasmas, Fluids, and Related Interdisciplinary Topics.* 50:1579-1588.

Fygenson, D. K., H. Flyvbjerg, K. Sneppen, A. Libchaber, and S. Leibler. 1995. Spontaneous Nucleation of Microtubules. *Physical Review E.* 51:5058-5063.

Gildersleeve, R. F., A. R. Cross, K. E. Cullen, A. P. Fagen, and R. C. Williams, Jr. 1992. Microtubules grow and shorten at intrinsically variable rates. *J Biol Chem.* 267:7995-8006.

Guenette, S., and F. Solomon. 1993. Microtubule assembly: Pathways, dynamics, and regulators. *Current Opinion in Structural Biology.* 3:198-201.

Hoh, J. H. 1998. Functional protein domains from the thermally driven motion of polypeptide chains: a proposal. *Proteins.* 32:223-228.

Howard, W. D., and S. N. Timasheff. 1986. GDP state of tubulin: stabilization of double rings. *Biochemistry.* 25:8292-8300.

Hyman, A. A., D. Chretien, I. Arnal, and R. H. Wade. 1995. Structural changes accompanying GTP hydrolysis in microtubules: information from a slowly hydrolyzable analogue guanylyl-(alpha,beta)- methylene-diphosphonate. *J Cell Biol.* 128:117-125.

Hyman, A. A., S. Salser, D. N. Drechsel, N. Unwin, and T. J. Mitchison. 1992. Role of GTP hydrolysis in microtubule dynamics: information from a slowly hydrolyzable analogue, GMPCPP. *Mol Biol Cell.* 3:1155-1167.

Kirschner, M., and T. Mitchison. 1986. Beyond self-assembly: from microtubules to morphogenesis. *Cell.* 45:329-342.

Kresheck, G. C., H. Schneider, and H. A. Scheraga. 1965. The effect of D2-O on the thermal stability of proteins. Thermodynamic parameters for the transfer of model compounds from H2-O to D2-O. *J Phys Chem.* 69:3132-3144.

Mandelkow, E. M., E. Mandelkow, and R. A. Milligan. 1991. Microtubule dynamics and microtubule caps: a time-resolved cryo- electron microscopy study. *J Cell Biol.* 114:977-991.

Mandelkow, E. M., R. Schultheiss, R. Rapp, M. Muller, and E. Mandelkow. 1986. On the surface lattice of microtubules: helix starts, protofilament number, seam, and handedness. *J Cell Biol.* 102:1067-1073.

Martin, S. R., M. J. Schilstra, and P. M. Bayley. 1993. Dynamic instability of microtubules: Monte Carlo simulation and application to different types of microtubule lattice [see comments]. *Biophys J.* 65:578-596.

Melki, R., M. F. Carlier, D. Pantaloni, and S. N. Timasheff. 1989. Cold depolymerization of microtubules to double rings: geometric stabilization of assemblies. *Biochemistry.* 28:9143-9152.

Milner, S. T. 1991. Polymer Brushes. *Science.* 251:905-914.

Mitchison, T., and M. Kirschner. 1984. Dynamic instability of microtubule growth. *Nature.* 312:237-242.

Muller-Reichert, T., D. Chretien, F. Severin, and A. A. Hyman. 1998. Structural changes at microtubule ends accompanying GTP hydrolysis: information from a slowly hydrolyzable analogue of GTP, guanylyl (alpha,beta)methylenediphosphonate. *Proc Natl Acad Sci U S A.* 95:3661-3666.

Nicholson, W. V., M. Lee, K. H. Downing, and E. Nogales. 1999. Cryo-electron microscopy of GDP-tubulin rings. *Cell Biochem Biophys.* 31:175-183.

Nogales, E., K. H. Downing, L. A. Amos, and J. Lowe. 1998a. Tubulin and FtsZ form a distinct family of GTPases. *Nat Struct Biol.* 5:451-458.

Nogales, E., S. G. Wolf, and K. H. Downing. 1998b. Structure of the alpha beta tubulin dimer by electron crystallography [see comments] [published erratum appears in Nature 1998 May 14;393(6681):191]. *Nature.* 391:199-203.

Panda, D., G. Chakrabarti, J. Hudson, K. Pigg, H. P. Miller, L. Wilson, and R. H. Himes. 2000. Suppression of microtubule dynamic instability and treadmilling by deuterium oxide. *Biochemistry.* 39:5075-5081.

Priev, A., A. Almagor, S. Yedgar, and B. Gavish. 1996. Glycerol decreases the volume and compressibility of protein interior. *Biochemistry.* 35:2061-2066.

Ray, S., E. Meyhofer, R. A. Milligan, and J. Howard. 1993. Kinesin follows the microtubule's protofilament axis. *J Cell Biol.* 121:1083-1093.

Sage, C. R., C. A. Dougherty, A. S. Davis, R. G. Burns, L. Wilson, and K. W. Farrell. 1995. Site-directed mutagenesis of putative GTP-binding sites of yeast beta- tubulin: evidence that alpha-, beta-, and gamma-tubulins are atypical GTPases. *Biochemistry.* 34:16870.

Schilstra, M. J., P. M. Bayley, and S. R. Martin. 1991. The effect of solution composition on microtubule dynamic instability. *Biochem J.* 277:839-847.

Stewart, R. J., K. W. Farrell, and L. Wilson. 1990. Role of GTP hydrolysis in microtubule polymerization: evidence for a coupled hydrolysis mechanism. *Biochemistry.* 29:6489-6498.

Timasheff, S. N. 1993. The control of protein stability and association by weak interactions with water: how do solvents affect these processes? *Annu Rev Biophys Biomol Struct.* 22:67-97.

Tran, P. T., P. Joshi, and E. D. Salmon. 1997. How tubulin subunits are lost from the shortening ends of microtubules. *J Struct Biol.* 118:107-118.

Voter, W. A., E. T. O'Brien, and H. P. Erickson. 1991. Dilution-induced disassembly of microtubules: relation to dynamic instability and the GTP cap. *Cell Motil Cytoskeleton.* 18:55-62.

Vulevic, B., and J. J. Correia. 1997. Thermodynamic and structural analysis of microtubule assembly: the role of GTP hydrolysis. *Biophys J.* 72:1357-1375.

Wade, R. H., D. Chretien, and D. Job. 1990. Characterization of microtubule protofilament numbers. How does the surface lattice accommodate? *J Mol Biol.* 212:775-786.

Walker, R. A., N. K. Pryer, and E. D. Salmon. 1991. Dilution of individual microtubules observed in real time in vitro: evidence that cap size is small and independent of elongation rate. *J Cell Biol.* 114:73-81.

Weisenberg, R. C., W. J. Deery, and P. J. Dickinson. 1976. Tubulin-nucleotide interactions during the polymerization and depolymerization of microtubules. *Biochemistry.* 15:4248-4254.

Wilson, L., D. Panda, and M. A. Jordan. 1999. Modulation of microtubule dynamics by drugs: A paradigm for the actions of cellular regulators. *CELL STRUCTURE AND FUNCTION.* 24:329-335.